# Stacking-dependent shear modes in trilayer graphene

Chun Hung Lui,[1] Zhipeng Ye,[2] Courtney Keiser,[2] Eduardo B. Barros,[3] Rui He[2, *]

[1] *Department of Physics, Massachusetts Institute of Technology, Cambridge, Massachusetts 02139, USA*
[2] *Department of Physics, University of Northern Iowa, Cedar Falls, Iowa 50614, USA*
[3] *Departmento de Física, Universidade Federal do Ceará, Fortaleza, Ceará 60455-760, Brazil*

*Corresponding author (email: rui.he@uni.edu)

**Abstract:** We observed distinct interlayer shear mode Raman spectra for trilayer graphene with ABA and ABC stacking order. There are two rigid-plane shear-mode phonon branches in trilayer graphene. We found that ABA trilayers exhibit pronounced Raman response from the high-frequency shear branch, without any noticeable response from the low-frequency branch. In contrast, ABC trilayers exhibit no response from the high-frequency shear branch, but significant Raman response from the low-frequency branch. Such complementary behaviors of Raman shear modes can be explained by the distinct symmetry of the two trilayer allotropes. The strong stacking-order dependence was not found in the layer-breathing modes, and thus represents a unique characteristic of the shear modes.

Layer stacking sequence is crucial in defining the physical properties of few-layer graphene (FLG). For instance, twisted bilayer graphene exhibits distinct electronic and optical properties from bilayers with the common Bernal stacking [1-10]. Graphene trilayers with Bernal (ABA) and rhombohedral (ABC) stacking (Fig. 1) have considerably different electronic structure, infrared absorption, band-gap tunability, quantum Hall effects and many-body physics [11-18]. While the stacking order strongly affects the electronic properties of FLG, its influence on the phonons is expected to be small because the interlayer coupling is mediated by weak van der Waals force and the next-nearest-layer *lattice* coupling is negligible [19, 20]. Indeed, theoretical calculations show almost identical phonon energy band structure for ABA and ABC trilayer graphene [Fig. 1(c)] [21]. Previous experiments have reported stacking-sensitive Raman and infrared response of phonons in FLG [4-10, 22-27], but all these phenomena arise from their coupling to the electronic system, either through double-resonance Raman processes or interactions with the interband electronic transitions. The apparent stacking dependence of the phonon Raman modes (e.g. the 2D, LOZO' and R modes) and infrared absorption are only indirect manifestations of the underlying stacking-dependent electronic structure. Without the involvement of electronic structure, the intrinsic phonon properties have not been shown to be sensitive to the stacking sequence.

In contrast to such an expectation of weak stacking dependence for phonons, we show here that the interlayer shear mode phonons of trilayer graphene exhibit intrinsic Raman response that depends directly and dramatically on the stacking order of the graphene layers. These shear mode vibrations are generated by interlayer interactions [19, 20, 28]. They consist of rigid lateral displacement of adjacent graphene layers, with a total of two branches (high and low shear branches) in a trilayer system [Fig. 1(b-c)]. Due to their low energy (< 6 meV), they have only been observed recently in FLG [29-34] and other atomically thin layered crystals [35-39]. In this Letter, we carried out Raman measurements of the shear modes in trilayer graphene with ABA and ABC stacking order [Fig. 1(a)]. We observed the shear mode that belongs to the high shear branch (C mode) in ABA trilayers, as reported in previous studies [29-34]. This Raman shear mode was, however, not observed in ABC trilayers. Instead, we found a different Raman shear mode in ABC trilayers, with frequency corresponding to the low shear branch (C' mode). We therefore observed the high (low) shear mode exclusively in the ABA (ABC) stacking order. The distinct Raman response of the two trilayer allotropes originates from their different crystal symmetry: ABA trilayers have mirror symmetry, whereas ABC trilayers have inversion symmetry [Fig. 1(a)]. As a consequence, the shear modes have distinct symmetry and Raman activity, though with similar frequency, for the two types of trilayers. Interestingly, such a strong sensitivity to the stacking order is not shared by the other type of interlayer vibrations, the layer-breathing modes (LBMs) [Fig. 1(b-c)] [20, 24, 25, 28, 40-45], and is therefore a unique characteristic of the shear modes. By considering the symmetry and the atomic displacements of the normal modes, we are able to explain all the key findings in our experiment.

To probe the intrinsic property of the interlayer vibrations, we investigated free-standing trilayer graphene samples exfoliated from kish graphite on quartz substrates with pre-patterned trenches (4 μm width, 1 μm depth). The suspended sample areas over the trenches were isolated from any disturbance of the substrate. The layer number and stacking order were determined by optical contrast [inset of Fig. 2(a)], infrared absorption (see Supplemental Material) [12, 46, 47], and different line shape of the 2D Raman mode for ABA and ABC trilayers [Fig. 2(b)] [22-24]. We performed Raman measurements at room temperature with a commercial Horiba Labram micro-Raman system equipped with a 532-nm excitation laser, a 100× objective lens, a 1800-groove/mm grating, and a thermo-electric cooled charge-coupled device (CCD). The Raman setup provided access to frequencies down to 10 $cm^{-1}$, with spectral resolution 0.5 $cm^{-1}$. We made the measurements in argon-purged environment to reduce the background signal from the air molecules. The excitation laser was focused onto the samples with spot diameter < 1 μm, which is much smaller than the width of the trenches. High incident laser power (~9 mW) was used in the experiment. The local graphene temperature reached ~900 K as estimated from the ratio of anti-Stokes and Stokes G Raman modes [47]. The strong laser heating was found to be favorable to reveal the weak interlayer modes, especially the LBMs, because of the increase of phonon population and removal of surface adsorbates (Detailed study of the temperature effect can be found in Ref. [48]). A slight redshift of the mode frequency (~3 $cm^{-1}$) occurred due to the rise of graphene temperature. No degradation of the samples was found as revealed by the negligible D Raman band [Fig. 2(a)]. This is consistent with prior studies, which show that the stacking order of trilayer graphene is stable up to at least $T = 1100$ K in the argon environment [22].

Fig. 2(c) displays the baseline-corrected Raman spectra of suspended ABA and ABC trilayer graphene in the frequency range of 15 – 100 $cm^{-1}$ [47]. The spectra were measured at the same experimental conditions and normalized with the integration time. We observed two main peaks in each spectrum. These peaks were absent in suspended monolayer graphene, indicating that they arise from the interlayer vibrations. In particular, both the ABA and ABC spectra exhibit a sharp Raman peak at 57 cm$^{-1}$, which matches the frequency (~59 cm$^{-1}$) of the low-frequency LBM branch in the trilayer graphene phonon band structure calculated by J. A. Yan *et. al.* using density functional theory (DFT) [Fig. 1(c)] [21] and also in Refs. [19, 20, 28, 41]. The similar frequency and intensity of the LBM Raman peak in the ABA and ABC samples imply that the two stacking orders have the same interlayer coupling strength. This is reasonable because ABA and ABC trilayers have the same nearest-layer coupling configuration and the next-nearest-layer *lattice* coupling is small [19].

In the lower frequency region, however, we observed distinct Raman spectra for the two types of trilayers [Fig. 2(c)]. The ABA spectrum exhibits a sharp Raman peak at 33 cm$^{-1}$. This peak corresponds to the high-frequency shear mode (C mode) in trilayer graphene, as reported by previous studies [Fig. 1(c)] [29-34]. The Raman response of the low shear branch was unobservable. In contrast, the ABC spectrum does not show any noticeable Raman feature at 33 cm$^{-1}$, but instead, exhibits a pronounced peak at 19 cm$^{-1}$. Since the interlayer coupling strength is the same for both trilayer samples, as implied by their similar LBMs, we expect the same frequency for the corresponding shear branches in the ABA and ABC trilayers. The peak at 19 cm$^{-1}$ therefore should not correspond to the high shear mode in the ABC trilayer. Its frequency actually matches that of the low shear mode predicted by the linear-chain model and DFT calculations [Fig. 1(c)] [29], and it also overlaps with the frequency of the low shear mode measured in folded trilayer graphene [33, 34]. This 19 cm$^{-1}$ mode is thus attributed to the low shear mode (C' mode) in ABC trilayer graphene. Therefore, the high and low shear modes are observed exclusively in the ABA and ABC structure, respectively.

We can interpret the different Raman response of the ABA and ABC shear modes from symmetry analysis based on group theory (Fig. 1) [19, 28, 38, 49, 50]. The observation of a vibrational mode through Raman scattering depends on the symmetry selection rules and the scattering geometry. The Raman intensity is proportional to $|e_i \cdot \widetilde{R} \cdot e_s|^2$, where $\widetilde{R}$ is the Raman tensor, and $e_i$ and $e_s$ are the polarization vectors of the incident and scattered light, respectively.

The ABA trilayer graphene has the D$_{3h}$ symmetry group, with 18 normal vibrational modes at the Γ point: 2A$_1$'+4E'+4A$_2$''+2E''. They include two rigid-layer shear modes, each with double degeneracy [Fig. 1(c)]. The high shear (C) mode has E' symmetry with even parity under mirror reflection. The low shear (C') mode has E'' symmetry with odd parity under mirror reflection. Although both shear modes are Raman active, their Raman tensors have different forms given by group theory:

$$\widetilde{R}_{E'} = \begin{pmatrix} a & b & 0 \\ b & -a & 0 \\ 0 & 0 & 0 \end{pmatrix}, \quad \widetilde{R}_{E''} = \begin{pmatrix} 0 & 0 & c \\ 0 & 0 & d \\ c & d & 0 \end{pmatrix}. \tag{1}$$

These Raman tensors indicate that the Raman intensity of the E' and E" modes is finite only for the in-plane ($x$, $y$) and out-of-plane ($z$) polarization, respectively. In our experiment, we excited the graphene samples with normally incident laser and collected the scattered light by back-scattering geometry. The polarizations of both the incident and scattered light are parallel to the graphene plane ($x$, $y$). In this configuration, the Raman intensity of C mode with E' symmetry is finite, but the intensity of C' mode with E" symmetry is zero.

In contrast, ABC trilayer graphene has the $D_{3d}$ symmetry group, with 18 normal modes at the Γ point: $3A_{1g}+3E_g+3A_{2u}+3E_u$. The high shear (C) mode has $E_u$ symmetry with odd parity under inversion. It is Raman forbidden. The low shear (C') mode has $E_g$ symmetry with even parity under inversion. Its Raman tensor has the form:

$$\widetilde{R}_{E_g} = \begin{pmatrix} e & f & 0 \\ f & -e & 0 \\ 0 & 0 & 0 \end{pmatrix}. \qquad (2)$$

It is Raman active in our in-plane polarization configuration. Therefore, the observation of the high (low) shear mode and the absence of low (high) shear mode in ABA (ABC) trilayer graphene are imposed by the symmetry of the crystals.

We can also understand the stacking-dependent Raman behaviors intuitively from the atomic displacements of the shear modes. In a simple classical description, the magnitude of the Raman response of a vibrational mode is proportional to the rate of change of polarizability (α) in the normal coordinate (Q), i.e. $(\partial\alpha/\partial Q)_0$, evaluated at the equilibrium position Q = 0 [51]. Here Q corresponds to the overall layer displacement of the vibrations [Fig. 1(b)]. Fig. 3 displays the schematic change of the atomic positions in the unit cell of trilayer during the shear vibrational motions. For the C mode, the ABA structure exhibits distinct atomic configurations and hence, different polarizability α at Q < 0 and Q > 0. This leads to non-zero derivative of polarizability $(\partial\alpha/\partial Q)_0$ and hence finite Raman response. In contrast, the ABC structure exhibits rather similar atomic configurations at Q < 0 and Q > 0, which are identical to each other by inversion. Consequently, $(\partial\alpha/\partial Q)_0$ and the Raman response are zero. The C' mode has the opposite stacking dependent behaviors from the C mode. The ABA C'-mode shows similar configurations at Q < 0 and Q > 0, which are identical to one another by mirror reflection. This leads to zero (or nearly zero) $(\partial\alpha/\partial Q)_0$, consistent with the zero Raman intensity predicted by group theory. Therefore, although the C' mode is Raman active in ABA trilayers, its Raman response is expected to be very weak, and indeed unobservable in our experiment. In contrast, the ABC C'-mode exhibits distinct configurations at Q < 0 and Q > 0 and hence finite Raman response. Therefore, the stacking order, even with only a slight shift of the top layer [Fig. 1(a)], has profound influence on the equilibrium lattice conditions, giving rise to distinct Raman activity of the shear mode vibrations.

The above picture can be quantified in an effective bond polarizability model [52, 53]. For the electron cloud bonding two adjacent graphene layers, e.g. layers *A* and *B* as denoted in Fig. 1(a), the change of bond polarizability can be approximated as $\Delta\alpha_{AB} = \alpha'_{AB}(q_A - q_B)$, where $\alpha'_{AB}$ is the differential bond polarizability and $q_A$ and $q_B$ are the layer displacement from the equilibrium position [Fig. 1(b)]. The total change of trilayer polarizability is a sum of the changes in the two interlayer bonds, i.e. $\Delta\alpha_{ABA} = \Delta\alpha_{AB} + \Delta\alpha_{A'B}$ and $\Delta\alpha_{ABC} = \Delta\alpha_{AB} + \Delta\alpha_{CB}$ [see layer notation at Fig. 1(a)]. For ABA trilayer, the mirror symmetry requires $\alpha'_{AB} = \alpha'_{A'B}$. But for ABC trilayer, the inversion symmetry requires $\alpha'_{AB} = -\alpha'_{CB}$ (Fig. 3). We therefore obtain $\Delta\alpha_{ABA} = \alpha'_{AB}[(q_A - q_B) + (q_{A'} - q_B)]$ and $\Delta\alpha_{ABC} = \alpha'_{AB}[(q_A - q_B) - (q_C - q_B)]$. The normalized layer displacements $(q_A, q_B, q_{A' \text{ or } C})$ are $(1, -2, 1)/\sqrt{6}$ and $(1, 0, -1)/\sqrt{2}$ for the high (C) and low (C') shear modes, respectively [Fig. 1(a-b)]. It follows straightforwardly that

$$\begin{cases} \Delta\alpha_{ABA}^{C'} = 0; \quad \Delta\alpha_{ABA}^{C} \propto \sqrt{6}; \\ \Delta\alpha_{ABC}^{C'} \propto \sqrt{2}; \quad \Delta\alpha_{ABC}^{C} = 0; \end{cases} \quad (3)$$

The Raman intensity can be estimated as $I \propto (1+n)|\Delta\alpha|^2$, where $n$ is the phonon population given by the Bose-Einstein distribution. The results confirm our qualitative picture in Fig. 3 and further predict the ratio of Raman intensity between C' and C modes to be $\frac{I_{ABC}^{C'}}{I_{ABA}^{C}} \propto \frac{1+n_{ABC}^{C'}}{1+n_{ABA}^{C}} \left|\frac{\Delta\alpha_{ABC}^{C'}}{\Delta\alpha_{ABA}^{C}}\right|^2 \approx 0.57$. This C'/C intensity ratio agrees well with our experimental value (~0.5) [Fig. 2(c)] and varies only slightly with temperature for $T > 300$ K.

It is interesting to compare the behaviors of the shear modes and the LBMs. There are two LBMs in trilayer graphene [Fig. 1(b)]. The high (low) LBM has odd (even) parity under mirror reflection for ABA trilayers or inversion for ABC trilayers. According to group-theory analysis, the high LBM is Raman forbidden and the low LBM is Raman active, for both types of trilayers. Fig. 4 displays the atomic positions of the trilayer unit cell during the LBM vibrational motions. For both the ABA and ABC structures, the high LBM exhibits similar atomic configurations at Q < 0 and Q > 0, which are identical to one another by mirror reflection (or inversion). This leads to zero derivative of polarizability $(\partial\alpha/\partial Q)_0$ and Raman response. But the low LBM exhibits distinct configurations at Q < 0 and Q > 0, leading to finite Raman response. In the effective bond polarizability model discussed above, the low LBM is predicted to have the same Raman intensity for both ABA and ABC trilayers, consistent with our observation [Fig. 2(c)]. Therefore, the LBM Raman behaviors are similar in both types of trilayer graphene, in contrast to the distinct behaviors found in the shear modes.

Finally, we comment on the line width of the interlayer modes. After correcting the instrumental broadening (0.5 cm$^{-1}$), the full-width-at-half-maxima (FWHM) for the ABC C'-mode, ABA C-mode, ABC and ABA LBMs are 1.2, 1.4, 1.8 and 2.3 cm$^{-1}$, respectively. The LBMs are broader than the shear modes. This is reasonable because the line width generally arises from the decay of phonons into other lower-energy phonons or electron-hole pairs. The LBM phonons, with relatively high energy, have more anharmonic decay channels than the shear-mode phonons. They are also expected to excite the electron-hole pairs more effectively than the lower-energy shear-mode phonons, because the electron-phonon couplings are more strongly suppressed in lower energies by the state filling effect due to charge doping and thermal excitations [54]. Further investigations are needed to better understand these phonon-phonon and electron-phonon couplings.

In conclusion, our experiment revealed strong dependence of the shear-mode Raman response on the stacking sequence in trilayer graphene. We observed pronounced Raman response of the high-frequency and low-frequency shear modes in ABA trilayer and ABC trilayer, respectively, whereas the other shear modes are suppressed correspondingly. Such distinct Raman response, not involving any double-resonance Raman process with the electronic structure, arises directly from the distinct symmetry of the two trilayer allotropes. Similar strong stacking-order dependence is expected for the shear modes in graphene with higher number of layers and more complicated stacking sequences, as well as in other atomically thin layered crystals, such as transition metal dichalcogenides. With the rapid advance of Raman detection techniques and the continuously improving efficiency, the shear mode may serve as an effective probe of the crystalline structure of diverse two-dimensional materials.

We thank L. M. Malard for helpful discussion and J.- A. Yan for providing us with the DFT calculation of phonon band structure of trilayer graphene. Acknowledgment is made to the Donors of the American Chemical Society Petroleum Research Fund (Grant 53401-UNI10) for support of this research. R. H. also acknowledges support from UNI Faculty Summer Fellowship. E. B. B. acknowledges support from CNPq and Funcap.

[1] G. Li, A. Luican, J. M. B. Lopes dos Santos, A. H. Castro Neto, A. Reina, J. Kong, and E. Y. Andrei, Nat. Phys. **6**, 109-113 (2010).
[2] Y. Wang, Z. Ni, L. Liu, Y. Liu, C. Cong, T. Yu, X. Wang, D. Shen, and Z. Shen, ACS Nano **4**, 4074-4080 (2010).
[3] J. T. Robinson, S. W. Schmucker, C. B. Diaconescu, J. P. Long, J. C. Culbertson, T. Ohta, A. L. Friedman, and T. E. Beechem, ACS Nano **7**, 637-644 (2012).
[4] V. Carozo, C. M. Almeida, E. H. M. Ferreira, L. G. Cançado, C. A. Achete, and A. Jorio, Nano Lett. **11**, 4527-4534 (2011).
[5] K. Kim, S. Coh, L. Z. Tan, W. Regan, J. M. Yuk, E. Chatterjee, M. F. Crommie, M. L. Cohen, S. G. Louie, and A. Zettl, Phys. Rev. Lett. **108**, 246103 (2012).
[6] R. W. Havener, H. Zhuang, L. Brown, R. G. Hennig, and J. Park, Nano Lett. **12**, 3162-3167 (2012).
[7] J. Campos-Delgado, L. Cançado, C. Achete, A. Jorio, and J.-P. Raskin, Nano Res. **6**, 269-274 (2013).
[8] A. Righi, P. Venezuela, H. Chacham, S. D. Costa, C. Fantini, R. S. Ruoff, L. Colombo, W. S. Bacsa, and M. A. Pimenta, Solid State Commun. **175**, 13-17 (2013).
[9] A. Jorio, and L. G. Cancado, Solid State Commun. **175**, 3-12 (2013).
[10] R. He, T.-F. Chung, C. Delaney, C. Keiser, L. A. Jauregui, P. M. Shand, C. C. Chancey, Y. Wang, J. Bao, and Y. P. Chen, Nano Lett. **13**, 3594-3601 (2013).

[11] M. F. Craciun, S. Russo, M. Yamamoto, J. B. Oostinga, A. F. Morpurgo, and S. Thrucha, Nat. Nanotech. **4**, 383-388 (2009).
[12] C. H. Lui, Z. Li, K. F. Mak, E. Cappelluti, and T. F. Heinz, Nat. Phys. **7**, 944-947 (2011).
[13] W. Bao, L. Jing, J. Velasco, Y. Lee, G. Liu, D. Tran, B. Standley, M. Aykol, S. B. Cronin, D. Smirnov, M. Koshino, E. McCann, M. Bockrath, and C. N. Lau, Nat. Phys. **7**, 948-952 (2011).
[14] L. Zhang, Y. Zhang, J. Camacho, M. Khodas, and I. Zaliznyak, Nat. Phys. **7**, 953-957 (2011).
[15] A. Kumar, W. Escoffier, J. M. Poumirol, C. Faugeras, D. P. Arovas, M. M. Fogler, F. Guinea, S. Roche, M. Goiran, and B. Raquet, Phys. Rev. Lett. **107**, 126806 (2011).
[16] S. H. Jhang, M. F. Craciun, S. Schmidmeier, S. Tokumitsu, S. Russo, M. Yamamoto, Y. Skourski, J. Wosnitza, S. Tarucha, J. Eroms, and C. Strunk, Phys. Rev. B **84**, 161408 (2011).
[17] T. Taychatanapat, K. Watanabe, T. Taniguchi, and P. Jarillo-Herrero, Nat. Phys. **7**, 621-625 (2011).
[18] K. Zou, F. Zhang, C. Clapp, A. H. MacDonald, and J. Zhu, Nano Lett. **13**, 369-373 (2013).
[19] H. Wang, Y. Wang, X. Cao, M. Feng, and G. Lan, J. Raman Spectrosc. **40**, 1791-1796 (2009).
[20] K. H. Michel, and B. Verberck, Phys. Rev. B **85**, 094303 (2012).
[21] J. A. Yan, W. Y. Ruan, and M. Y. Chou, Phys. Rev. B *77*, 125401 (2008), and private communication.
[22] C. H. Lui, Z. Li, Z. Chen, P. V. Klimov, L. E. Brus, and T. F. Heinz, Nano Lett. **11**, 164-169 (2010).
[23] C. Cong, T. Yu, K. Sato, J. Shang, R. Saito, G. F. Dresselhaus, and M. S. Dresselhaus, ACS Nano **5**, 8760-8768 (2011).
[24] C. H. Lui, L. M. Malard, S. Kim, G. Lantz, F. E. Laverge, R. Saito, and T. F. Heinz, Nano Lett. **12**, 5539-5544 (2012).
[25] C. H. Lui, and T. F. Heinz, Phys. Rev. B **87**, 121404(R) (2013).
[26] Z. Li, C. H. Lui, E. Cappelluti, L. Benfatto, K. F. Mak, G. L. Carr, J. Shan, and T. F. Heinz, Phys. Rev. Lett. **108**, 156801 (2012).
[27] C. H. Lui, E. Cappelluti, Z. Li, and T. F. Heinz, Phys. Rev. Lett. **110**, 185504 (2013).
[28] S. K. Saha, U. V. Waghmare, H. R. Krishnamurthy, and A. K. Sood, Phys. Rev. B **78**, 165421 (2008).
[29] P. H. Tan, W. P. Han, W. J. Zhao, Z. H. Wu, K. Chang, H. Wang, Y. F. Wang, N. Bonini, N. Marzari, N. Pugno, G. Savini, A. Lombardo, and A. C. Ferrari, Nat. Mater. **11**, 294-300 (2012).
[30] D. Boschetto, L. Malard, C. H. Lui, K. F. Mak, Z. Li, H. Yan, and T. F. Heinz, Nano Lett. **13**, 4620-4623 (2013).
[31] J. Tsurumi, Y. Saito, and P. Verma, Chem. Phys. Lett. **557**, 114-117 (2013).

[32] A. C. Ferrari, and D. M. Basko, Nat. Nanotech. **8**, 235-246 (2013).
[33] C. C. Cong, and T. Y. Yu, arXiv:1312.6928 (2014).
[34] P. H. Tan, J. B. Wu, W.-P. Han, W.-J. Zhao, X. Zhang, H. Wang, and Y.-F. Wang, arXiv:1401.0804 (2014).
[35] G. Plechinger, S. Heydrich, J. Eroms, D. Weiss, C. Schüller, and T. Korn, Appl. Phys. Lett. **101**, 101906 (2012).
[36] H. Zeng, B. Zhu, K. Liu, J. Fan, X. Cui, and Q. M. Zhang, Phys. Rev. B **86**, 241301(R) (2012).
[37] X. Zhang, W. P. Han, J. B. Wu, S. Milana, Y. Lu, Q. Q. Li, A. C. Ferrari, and P. H. Tan, Phys. Rev. B **87**, 115413 (2013).
[38] Y. Zhao, X. Luo, H. Li, J. Zhang, P. T. Araujo, C. K. Gan, J. Wu, H. Zhang, S. Y. Quek, M. S. Dresselhaus, and Q. Xiong, Nano Lett. **13**, 1007-1015 (2013).
[39] M. Boukhicha, M. Calandra, M.-A. Measson, O. Lancry, and A. Shukla, Phys. Rev. B **87**, 195316 (2013).
[40] F. Herziger, P. May, and J. Maultzsch, Phys. Rev. B **85**, 235447 (2012).
[41] L. J. Karssemeijer, and A. Fasolino, Surface Science **605**, 1611-1615 (2011).
[42] S. Kitipornchai, X. Q. He, and K. M. Liew, Phys. Rev. B **72**, 075443 (2005).
[43] I. V. Lebedeva, A. A. Knizhnik, A. M. Popov, Y. E. Lozovik, and B. V. Potapkin, Phys. Chem. Chem. Phys. **13**, 5687-5695 (2011).
[44] K. Sato, J. S. Park, R. Saito, C. Cong, T. Yu, C. H. Lui, T. F. Heinz, G. Dresselhaus, and M. S. Dresselhaus, Phys. Rev. B **84**, 035419 (2011).
[45] P. T. Araujo, D. L. Mafra, K. Sato, R. Saito, J. Kong, and M. S. Dresselhaus, Sci. Rep. **2**, 1017 (2012).
[46] K. F. Mak, M. Y. Sfeir, J. A. Misewich, and T. F. Heinz, Proc. Natl. Acad. Sci. U.S.A. **107**, 14999-15004 (2010).
[47] See Supplemental Material for infrared spectra, original low-frequency Raman spectra, anti-Stokes and Stokes G mode spectra of ABA and ABC trilayer graphene.
[48] C. H. Lui, Z. Ye, C. Keiser, X. Xiao, and R. He, arXiv:1405.6735, (2014).
[49] L. M. Malard, M. H. D. Guimarães, D. L. Mafra, M. S. C. Mazzoni, and A. Jorio, Phys. Rev. B **79**, 125426 (2009).
[50] J.-W. Jiang, H. Tang, B.-S. Wang, and Z.-B. Su, Phys. Rev. B **77**, 235421 (2008).
[51] J. R. Ferraro, K. Nakamoto, and C. W. Brown, *Introductory Raman Spectroscopy*. ( (2nd ed., Academic Press, 2003)).
[52] D. W. Snoke, and M. Cardona, Solid State Commun. **87**, 121-126 (1993).
[53] D. Bermejo, S. Montero, M. Cardona, and A. Muramatsu, Solid State Commun. **42**, 153-155 (1982).
[54] J. Yan, Y. Zhang, P. Kim, and A. Pinczuk, Phys. Rev. Lett. **98**, 166802 (2007).

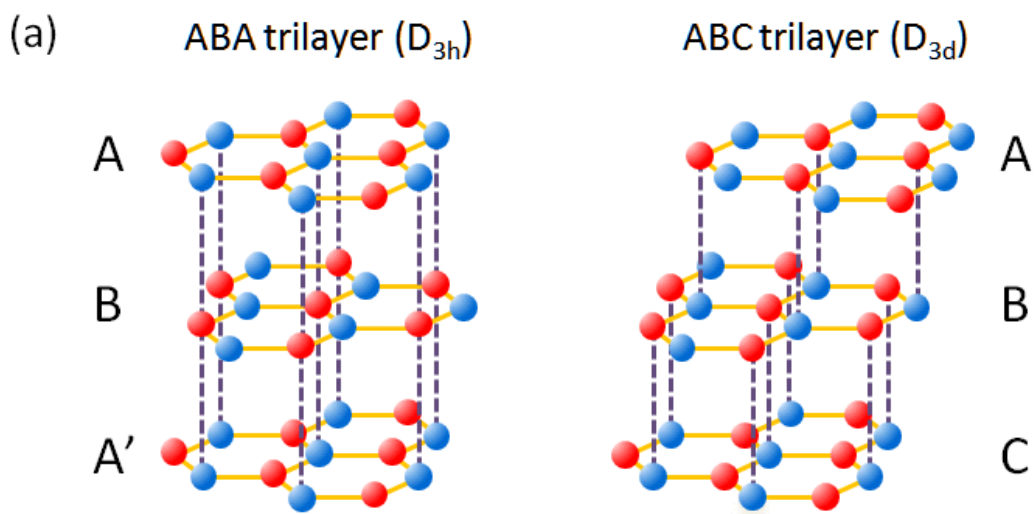

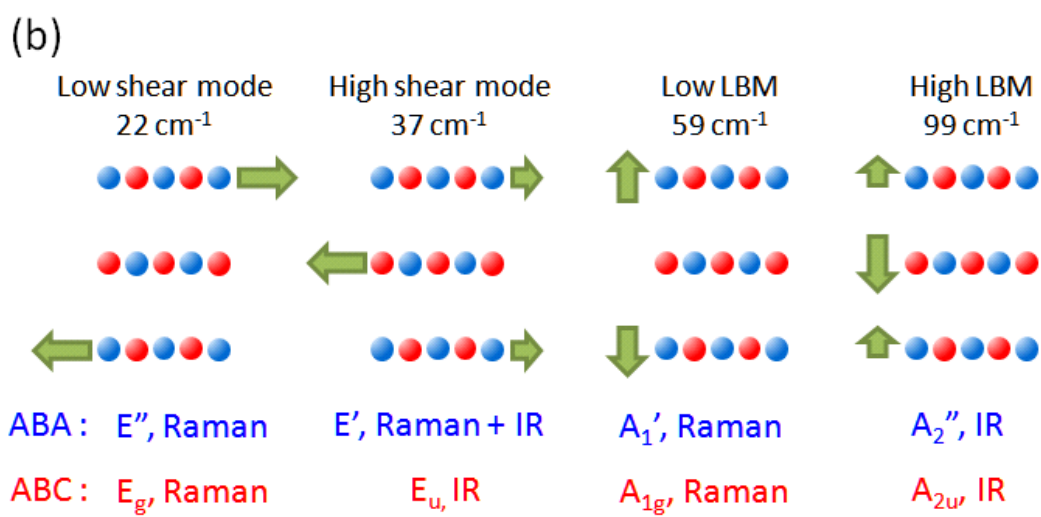

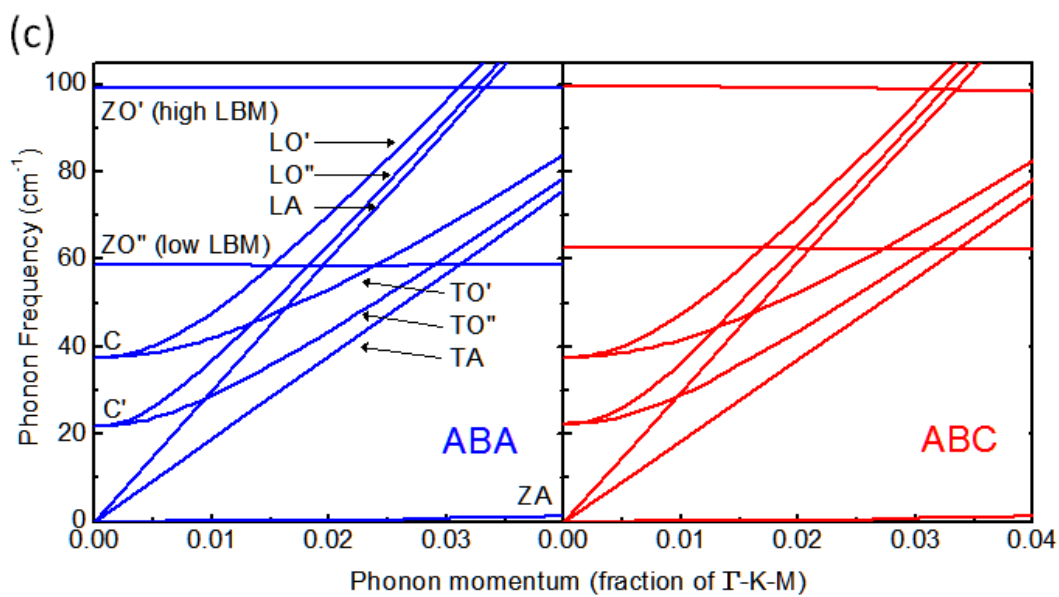

FIG. 1. (a) Crystal structure and symmetry group of trilayer graphene with ABA and ABC stacking order. The red and blue dots represent the two nonequivalent sublattices of the graphene honeycomb structure. The three layers are denoted as A, B and A' (or C). The ABC structure can be viewed as generated from the ABA structure by laterally shifting the top graphene layer by one carbon-carbon bond length. (b) Schematic representations of the atomic displacements for the rigid-plane interlayer vibrational modes. The arrows represent the direction and magnitude of the layer displacement. For each normal mode, we denote the corresponding predicted frequency from panel (c), symmetry, Raman and infrared (IR) activity for the two stacking orders. We note that the E" mode in ABA trilayer is unobservable in our measurement geometry, although this mode is Raman active (see the text for discussion). (c) Low-energy phonon band structure of ABA and ABC trilayer graphene calculated by density functional theory. The results are adapted from Ref. [21]. The high (ZO') and low (ZO") LBMs as well as the high (C) and low (C') shear modes at the $\Gamma$ point are denoted in the figure.

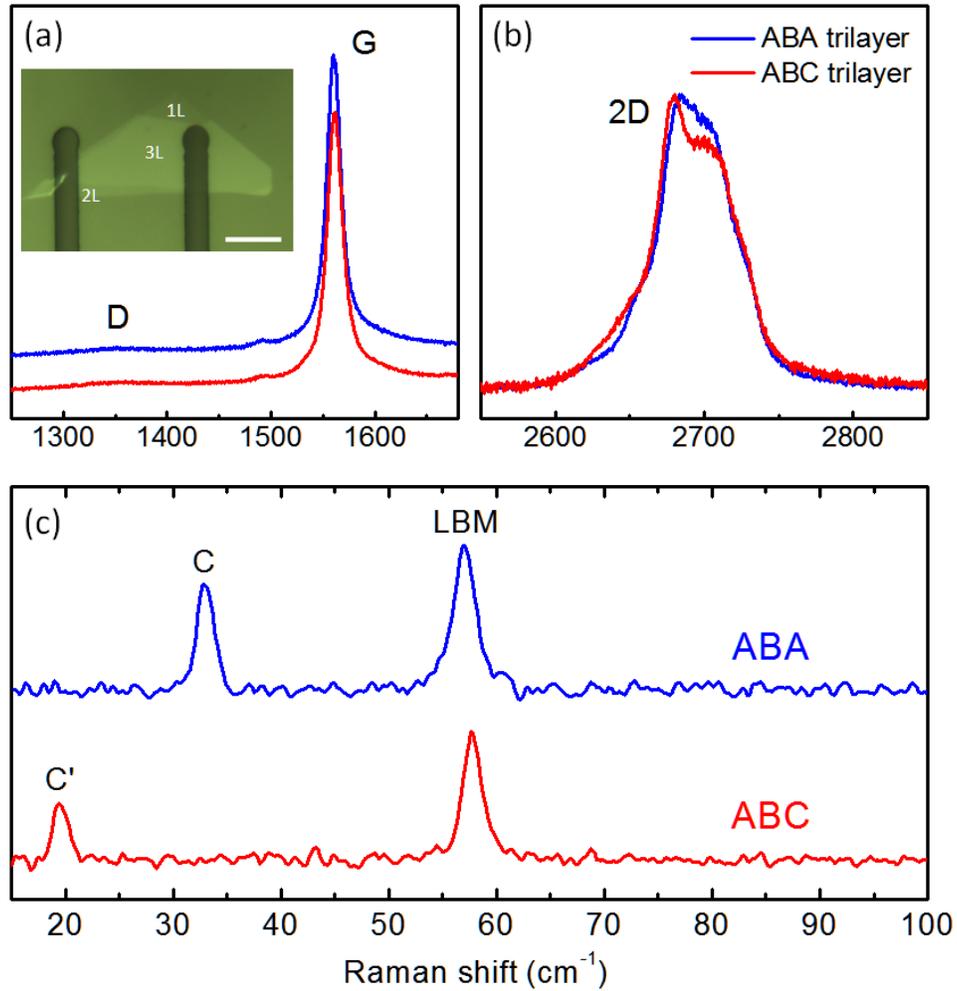

FIG. 2. (a) Raman spectra of D and G modes from suspended ABA (blue) and ABC (red) trilayer graphene with laser power ~ 9 mW. The negligible D band indicates the high quality of our samples under laser heating. The inset shows the optical image of an ABC graphene trilayer deposited on a quartz substrate with prep-patterned trenches. The scale bar is 10 μm. (b) Raman spectra of 2D mode for suspended ABA and ABC trilayer. Lower laser power (~1 mW) was used here to avoid thermal broadening of the Raman lines. (c) Low-frequency Raman spectra for suspended ABA and ABC trilayer with ~9 mW laser excitation. The broad background due to Rayleigh scattering has been subtracted. Original spectra are shown in the Supplemental Material. The Raman peaks corresponding to the layer-breathing mode (LBM), high- (C) and low (C') shear modes are denoted. All measurements were conducted at room temperature with a 532-nm laser excitation.

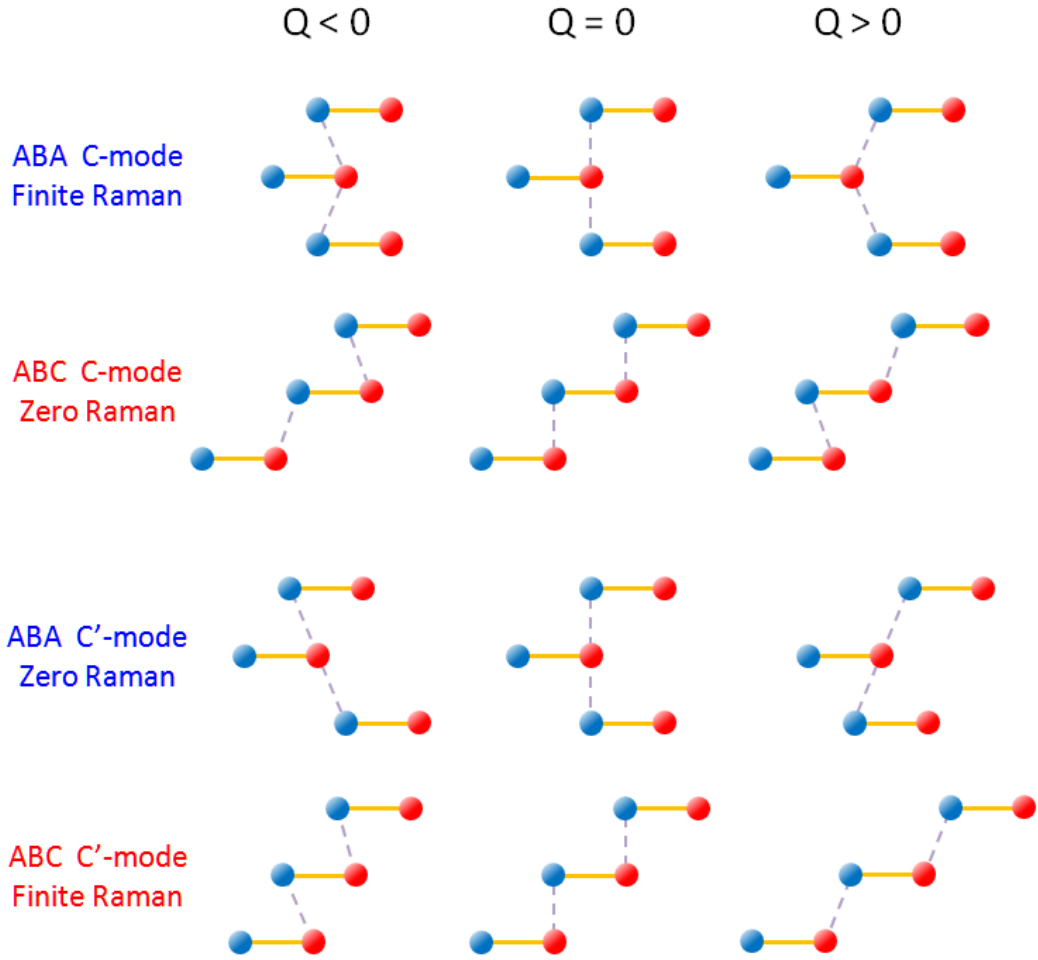

FIG. 3. Schematic representations of the change of atomic positions in the unit cell during the shear-mode vibrations for ABA and ABC trilayer graphene. The Raman activity is indicated for each normal mode. The red and blue dots represent the A and B carbon atoms in the trilayer unit cell. The yellow lines and purple dashed lines represent, respectively, the nearest-neighbor intralayer and interlayer coupling. From left to right, the columns represent the atomic positions of the normal modes at $Q < 0$, $Q = 0$ and $Q > 0$. Q is the overall layer displacement of the interlayer vibrations, which corresponds to the arrows in Fig. 1(b).

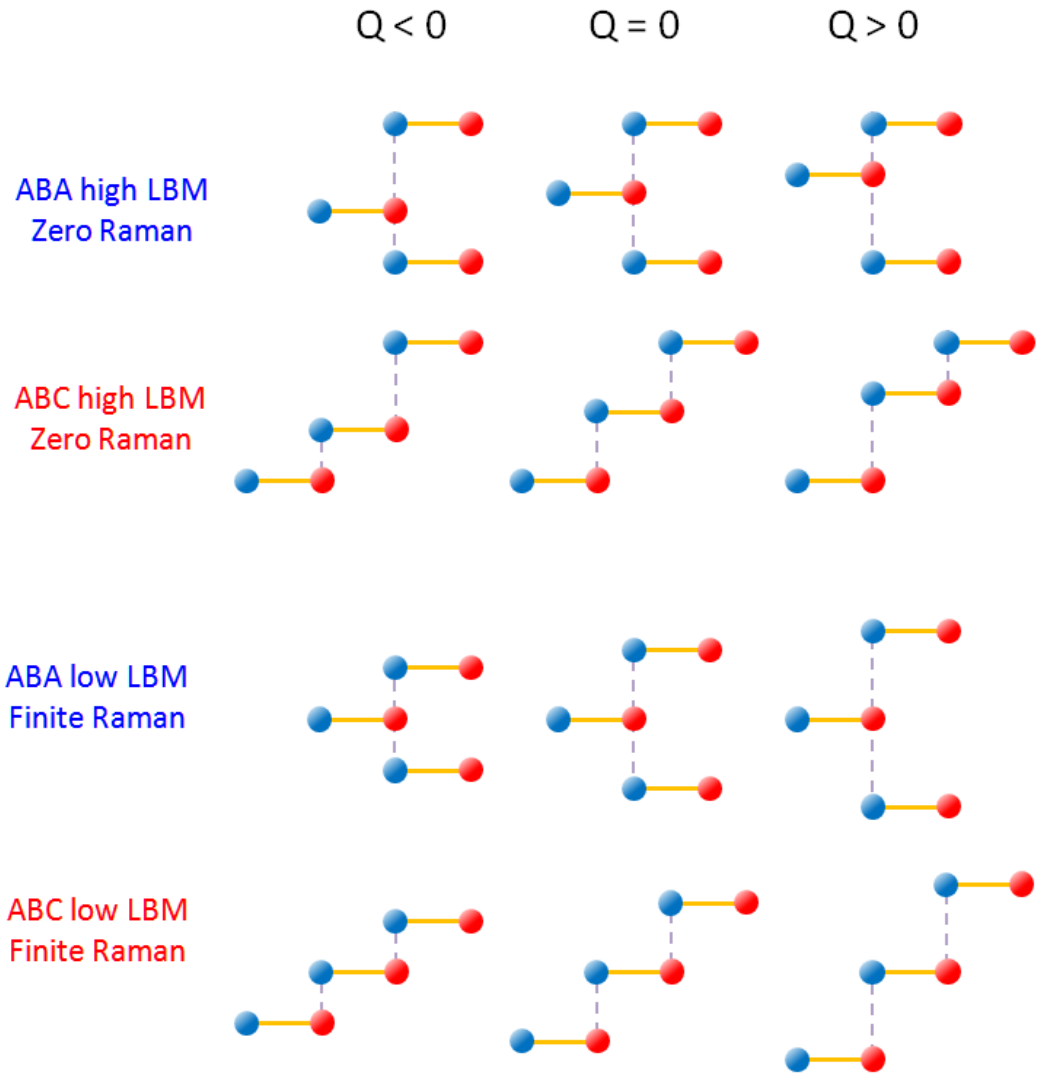

FIG. 4. Schematic representations of the atomic displacements as in Fig. 3, but for the layer breathing modes (LBMs) in trilayer graphene.

# *Supplemental Material of*
# Stacking-dependent shear modes in trilayer graphene

Chun Hung Lui,[1] Zhipeng Ye,[2] Courtney Keiser,[2] Eduardo B. Barros,[3] Rui He[2, *]

[1] *Department of Physics, Massachusetts Institute of Technology, Cambridge, Massachusetts 02139, USA*
[2] *Department of Physics, University of Northern Iowa, Cedar Falls, Iowa 50614, USA*
[3] *Departmento de Física, Universidade Federal do Ceará, Fortaleza, Ceará 60455-760, Brazil*

*Corresponding author (email: rui.he@uni.edu)

## 1. Determination of the layer number and stacking order by infrared spectroscopy

We exfoliated few-layer graphene (FLG) samples on quartz substrates with arrays of pre-patterned micron-scale trenches. Afterward, we determined the layer number and stacking order of the suspended FLG samples by measuring the infrared absorption spectrum at the adjacent FLG areas supported on the substrates. We performed the experiment with a micro-Fourier Transform Infrared spectrometer with a globar source and an HgCdTe detector. From the reflectance spectra of the FLG area on the substrate ($R_{FLG}$) and of the bare substrate ($R_{sub}$), we obtained the optical sheet conductivity $\sigma(\hbar\omega)$ from the following formula [S1] :

$$\delta_R = \frac{R_{FLG} - R_{sub}}{R_{sub}} = \frac{4}{n_{sub}^2 - 1} \frac{4\pi}{c} \sigma .\qquad(S1)$$

Here $c$ denotes the speed of light and $n_{sub}$ is the refractive index of the quartz substrate. Figure S1 displays the two types of absorption spectra found in trilayer graphene. At the high-energy range (> 0.7 eV), the optical conductivity is nearly independent of the stacking order, with each graphene layer contributing a value of $\pi e^2/2h$. We can readily identify the trilayer thickness from the expected conductivity value of $3\times\pi e^2/2h$ at $\hbar\omega >$ 0.7 eV. At the low-energy range (< 0.7 eV), however, ABA and ABC trilayers exhibit different absorption features due to their distinct electronic structure in this energy range. We could identify the ABA and ABC stacking order by their characteristic absorption band at $\hbar\omega$ ~0.52 eV and ~0.35 eV, respectively. We note that the double peaks at the ABC spectrum correspond to the induction of a band gap due to the unintentional charge doping of the substrate [S2]. Detailed analysis of these infrared absorption spectra can be found in the literature [S1-4].

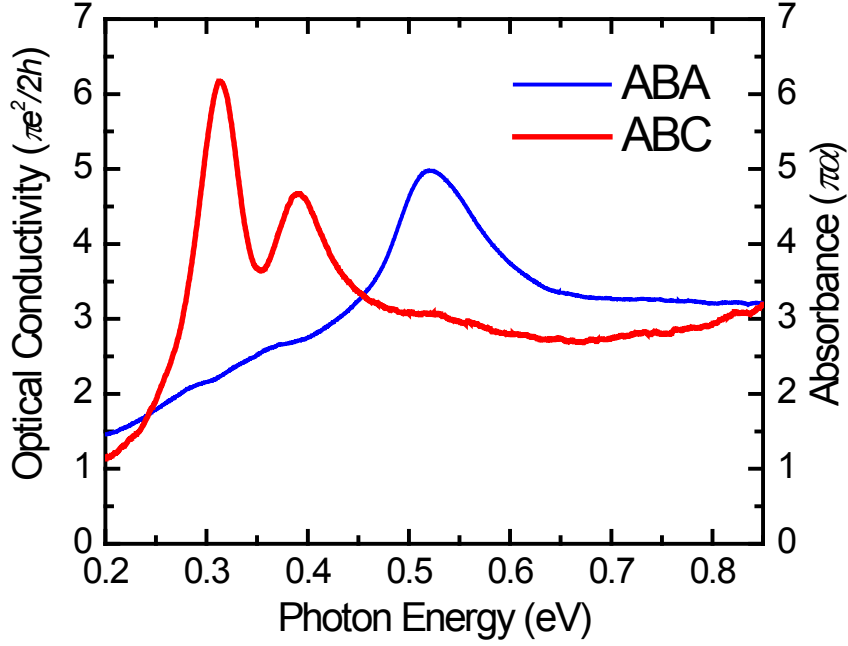

FIG. S1. Optical sheet conductivity spectra of trilayer graphene with ABA and ABC stacking order. We also show the corresponding absorbance in the right axis, which is proportional to the conductivity for atomically thin graphene samples.

## 2. Extraction of interlayer Raman modes in trilayer graphene

Fig. S2 displays the low-frequency Raman spectra of suspended ABA and ABC trilayer graphene in the initial stage of our analysis. The spectra exhibit a broad background due to the elastic scattering. We note that the noise level of our spectra is much lower than what appears to be. The small features in the spectra are not noise, but Raman signals from the residue air molecules in our setup. The interlayer modes are observed upon these background signals. Although the C' mode in ABC trilayer graphene is somewhat obscured by the relatively steep background, this mode can still be identified unambiguously. This feature is not found in ABA trilayer, bilayer and monolayer graphene. It was observed at all positions of ABC trilayer samples and absent when we removed the samples. In order to reveal the weak interlayer modes, we first subtracted the elastic scattering background by fitting the spectra without the main Raman features using a hybrid exponential and polynomial function. Afterward, by comparing with the air spectrum, we further removed the Raman signals from the air. The resultant baseline-corrected spectra are presented in Fig. 2(c) of the main paper.

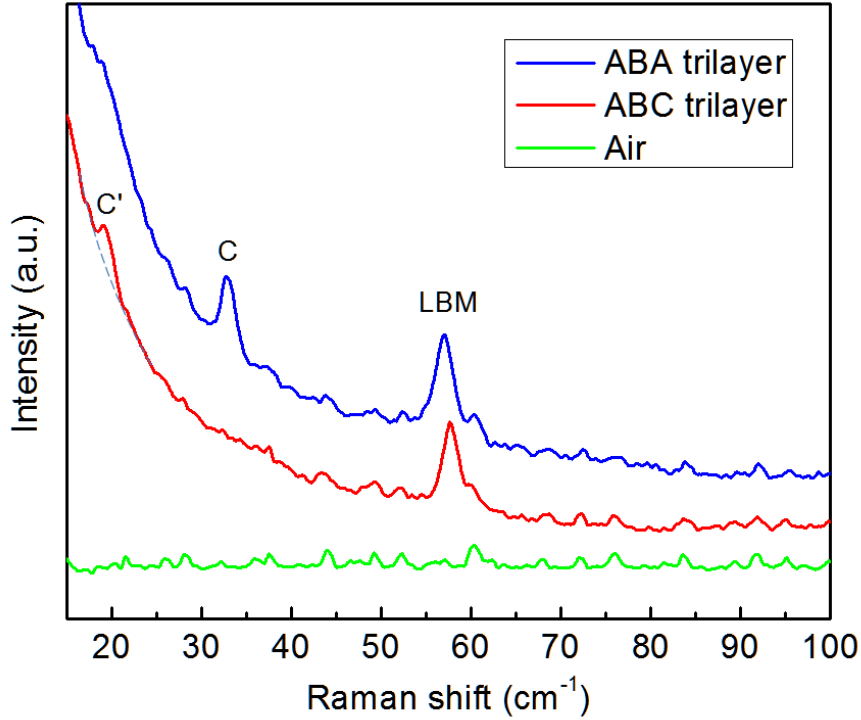

FIG. S2. Low-frequency Raman spectra of suspended ABA and ABC trilayer graphene, in comparison with a spectrum of residue air molecules in our argon-purged setup. The interlayer modes (C', C and LBM) are observed on a broad elastic scattering background with small Raman features from air molecules. The dashed line is a guide to the eye to show the C' mode. The measurement was made using a 532 nm laser with ~ 9 mW incident power.

## 3. Graphene temperature under laser excitation

We made use of high laser power in our experiment to reveal the weak interlayer Raman modes in trilayer graphene. To estimate the local graphene temperature due to laser heating, we have measured the anti-Stokes and Stokes G Raman mode spectra (Fig. S3) under the same condition that the low-frequency modes were measured. From the ratio of their intensities after taking into account our instrumental efficiency, we estimated the temperature to be $T \sim 900$ K. This temperature was used to calculate the shear mode phonon population and Raman intensity in the main paper. Detailed investigation of the influence of temperature on the interlayer modes can be found in our other publication [S5].

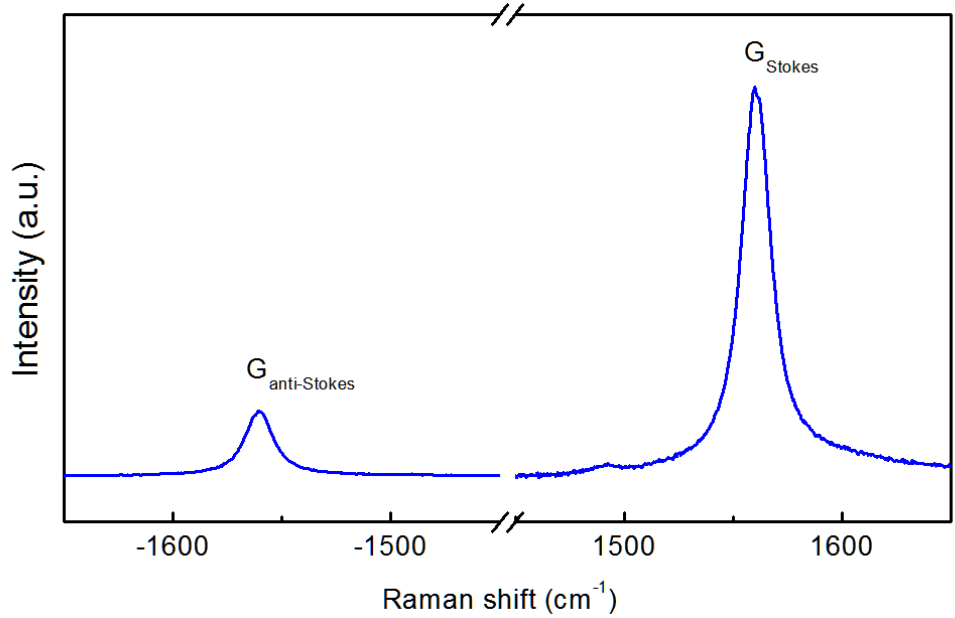

FIG. S3. Raman spectra of anti-Stokes and Stokes G modes for a suspended ABA trilayer graphene, under the same conditions of the measurement of low-frequency interlayer modes. The measurement was made using a 532-nm laser with ~9 mW excitation power. Similar spectra were observed for suspended ABC trilayer graphene.

**Supplemental References**

[S1] K. F. Mak, M. Y. Sfeir, Y. Wu, C. H. Lui, J. A. Misewich, and T. F. Heinz, Phys. Rev. Lett. **101**, 196405 (2008).
[S2] C. H. Lui, Z. Li, K. F. Mak, E. Cappelluti, and T. F. Heinz, Nat. Phys. **7**, 944-947 (2011).
[S3] K. F. Mak, J. Shan, and T. F. Heinz, Phys. Rev. Lett. **104**, 176404 (2010).
[S4] K. F. Mak, M. Y. Sfeir, J. A. Misewich, and T. F. Heinz, Proc. Natl. Acad. Sci. U.S.A. **107**, 14999-15004 (2010).
[S5] C. H. Lui, Z. Ye, C. Keiser, X. Xiao, and R. He, arXiv:1405.6735, (2014).